\documentclass[
    ,final            
  ]
  {aipproc}

\layoutstyle{8x11double}

\begin{document}

\title{Can the frequency dependent \textit{isobaric} specific heat be measured by thermal effusion methods?}

\classification{64.70.Pf}
\keywords      {glass transition, longitudinal specific heat, specific heat spectroscopy, thermal effusion.}

\author{T. Christensen, N. B. Olsen, J. C. Dyre.}{
  address={DNRF centre ``Glass and Time'', IMFUFA, Department of Sciences,\\ Roskilde University, Postbox 260, DK-4000 Roskilde, Denmark.}
}

\newcommand{\bu}{{\bf u}}
\newcommand{\br}{{\bf r}}
\newcommand{\bnul}{{\bf 0}}
\newcommand{\bnab}{{\bf\nabla}}
\newcommand{\pa}{\partial}
\newcommand{\dt}{\delta T}
\newcommand{\bv}{\beta_V}
\newcommand{\lap}{\nabla^2}
\newcommand{\divu}{\bnab\cdot\bu}

\newcommand{\tr}{\tilde r}
\newcommand{\bsig}{\bf \sigma}

\begin{abstract}
It has recently been shown that plane-plate heat effusion methods devised for wide-frequency specific-heat spectroscopy do not give the isobaric specific heat, but rather the so-called longitudinal specific heat. Here it is shown that heat effusion in a spherical symmetric geometry also involves the longitudinal specific heat. 
\end{abstract}

\maketitle


\section{Introduction}

The frequency-dependent specific heat is one of the most fundamental thermoviscoelastic response functions characterizing relaxation of liquid structure in highly viscous liquids. The measurement of this quantity as an alternative to enthalpy relaxation studies was conceived more than two decades ago \cite{bir85,chr85}.

When the specific heat is frequency-dependent one faces experimentally the problem of separating out the trivial frequency dependence from the slow propagation of heat. This can be solved in two ways. One can go to the thermally thin limit \cite{chr85,sch07} where the sample is small compared to the thermal diffusion length $l_D$. $l_D$ is inversely proportional to the square root of the frequency and typically $0.1$mm even at $1$Hz. So this condition can be difficult to fulfill over a wide frequency range unless the sample is very small \cite{sch07}.

The other - effusion - approach is to choose the sample size much larger than the thermal wavelength. This is the thermally thick limit and it is easier to realize over a wide frequency range \cite{bir86}. Both methods have to take due account of the stresses coming from the material supporting the sample associated with different thermal expansion coefficients. The effusion method suffers additionally from thermal stresses produced within the liquid itself. It is only the latter problem we consider in this paper. 

In the effusion methods one typically produces a harmonically varying heat current $\textnormal{Re} \lbrace P_\omega e^{i\omega t} \rbrace$ at a surface in contact with the liquid. The corresponding temperature response  $\textnormal{Re} \lbrace T_\omega e^{i\omega t} \rbrace$ on the very same surface is measured. Since the response is linear in the stimulus it is convenient to introduce the complex thermal impedance
\begin{equation}
Z=\frac{T_\omega}{P_\omega}
\end{equation}
The thermal impedance of a sample of volume $V$ and volume specific heat $c$ is
\begin{equation}
Z=\frac{1}{i\omega c V}
\end{equation}
in the thermally thin limit.

If on the other hand planar thermal waves effuses from a plate of area $A$ into a liquid the thermal impedance is
\begin{equation}\label{e0}
Z=\frac{1}{A \sqrt{i\omega c \lambda}}
\end{equation}
in the thermally thick limit \cite{bir85}. Here $\lambda$ is the thermal conductivity.

It has allways tacitly been assumed that measurements done at ambient  pressure are isobaric and that the $c$ of formula (\ref{e0}) is $c_p$. However it was formerly stated \cite{chr97} and recently shown \cite{chr07} that the ordinary heat diffusion with a complex diffusion constant does not describe the experimental situation adequately. For unidirectional heat effusion it was shown that the effective specific heat measured is the so-called longitudinal specific heat $c_l(\omega)$ which is between the isochoric, $c_V(\omega)$ and isobaric, $c_p(\omega)$ specific heats. Denoting the adiabatic and isothermal bulk moduli by $K_s(\omega)$ and $K_T(\omega)$ respectively and the shear modulus by $G(\omega)$ one can write the adiabatic and isothermal longitudinal moduli as $M_s(\omega)=K_s(\omega)+4/3 G(\omega)$ and  $M_T(\omega)=K_T(\omega)+4/3 G(\omega)$. Now $c_l(\omega)$ is related to $c_V(\omega)$ as \cite{chr07}
\begin{equation}\label{e1}
c_l(\omega)=\frac{M_s(\omega)}{M_T(\omega)} c_V(\omega),
\end{equation}
whereas $c_p(\omega)$ is related to $c_V(\omega)$ as 
\begin{equation}\label{e2}
c_p(\omega)=\frac{K_s(\omega)}{K_T(\omega)} c_V(\omega).
\end{equation}
In an easily flowing liquid $G(\omega)$ is negligible since $1/\omega$ is large compared to the Maxwell relaxation time, $\tau_M$, and there is no difference between $c_l$ and $c_p$. However in a highly viscous liquid near the dynamic glass transition the shear modulus becomes comparable to the bulk moduli and the diffence between $c_l(\omega)$ and $c_p(\omega)$ becomes significant.

We show below that the same is true for heat effusion in a spherical symmetric geometry. Here one also obtains $c_l(\omega)$. Furthermore in spherical geometry one can also get the heat conductivity $\lambda$ and thus get $c_l(\omega)$ absolutely. The planar unidirectional method in fact gives only the effusivity, $\sqrt{\lambda c_l(\omega)}$; that is $c_l(\omega)$ is determined only to within a proportionality constant.

\section{Thermal and mechanical coupling}

\subsection{The general equations}
One cannot treat the diffusion of heat independently of the associated creation of strains or stresses. Let the temperature field, $T(\br,t)$ be described in terms of the small deviation $\dt(\br,t)=T(\br,t)-T_0$ from a reference temperature $T_0$ and denote the displacement field by $\bu=\bu(\br,t)$. Dealing with relaxation is most conveniently done in the frequency domain. Thus time dependence of the fields is given by the factor $e^{st}$, $s=i\omega$. Considering only cases where inertia can be neglected the equations that couple temperature and displacement are \cite{chr07}

\begin{eqnarray}
M_T\bnab(\divu)-\bv\bnab\dt-G\bnab\times(\bnab\times\bu)=0\label{e3}\\
c_V s\dt+\bv T_0 s \divu -\lambda\lap\dt=0\label{e4}.
\end{eqnarray}
Here the isochoric pressure coefficient $\bv(\omega)$ is defined in the constitutive equation for the trace of the stress tensor $\sigma$
\begin{equation}\label{e5}
 \frac{1}{3} \textnormal{tr} \left(\sigma\right)=K_T \divu -\bv \dt.
\end{equation}

\subsection{The isobaric case}

If the trace of the stress tensor is constant in time then the term $\bv T_0 s\divu$ in equation (\ref{e4}) becomes $T_0 \bv^2 / K_T \dt$. Since $T_0 \bv^2/K_T =c_p-c_V$ equation (\ref{e4}) now becomes the ordinary heat diffusion equation
\begin{equation}\label{e6}
s \dt=D_p\lap\dt,
 \end{equation} 
decoupled from the displacement field and with a diffusion constant involving the isobaric specific heat
\begin{equation}\label{e7}
D_p=\frac{\lambda}{c_p}
\end{equation} 

It is usually assumed that thermal experiments on liquids with a completely or partially free surface will be at isobaric conditions. However this is only true if the shear modulus $G$ can be neglected compared to bulkmodulus $K_T$. This condition fails near the glass transition and the full coupled problem of equations (\ref{e3}) and (\ref{e4}) has to be considered.

\subsection{Radial heat effusion from a spherical surface into an infinite media}
We would like to show here that the inherent problem of measuring $c_p$ is not only confined to one-dimensional heat flow in the geometry considered in \cite{chr07} where the associated displacement field is forced to be longitudinal.

The longitudinal specific heat also emerges in the thermal impedance against effusion out from a sphere. In the spherically symmetric case $\bnab\times\bu$ vanishes. If we denote differentiation with respect to $r$ by a prime (\ref{e3}) and (\ref{e4}) becomes
\begin{eqnarray}\label{e8}
M_T(r^{-2}(r^2u)')'-\bv\dt'=0 \\ \label{e9}
c_v s \dt+ T_0 \bv s r^{-2}(r^2u)'-\lambda r^{-2}(r^2 \dt')'=0.
\end{eqnarray}
Define now the longitudinal specific heat,
\begin{equation}
 c_l \equiv c_V+T_0\frac{\bv^2}{M_T},
\end{equation}
the heat diffusion constant,
\begin{equation}
 D=\frac{\lambda}{c_l},
\end{equation}
and the wave vector
\begin{equation}
 k=\sqrt{\frac{s}{D}}.
\end{equation}
We thermally perturb the system by a harmonically varying heat current density $j_q e^{st}$ with $s=i\omega$ at the surface of radius $r_1$. If we impose the boundary conditions of vanishing fields at infinity and a hard core, $u(r_1)=0$ then the coupled solution is
\begin{eqnarray}
 \dt(r)=\frac{k}{s c_l}\frac{(kr_1)^2}{(1+kr_1)kr} e^{-k(r-r_1)} j_q\\
 u(r)=\frac{\bv}{M_T c_l s}(\frac{r_1}{r})^2(1-\frac{1+kr}{1+kr_1}e^{-k(r-r_1)} )j_q.
\end{eqnarray}
The total thermal impedance thus becomes
\begin{equation}\label{e15}
 Z \equiv \frac{\dt(r_1)}{4 \pi {r_1}^2 j_q}=\frac{1}{4 \pi \lambda r_1}\frac{1}{1+kr_1}
\end{equation}
or
\begin{equation}\label{e16}
 Z=\frac{1}{4 \pi \lambda r_1}\frac{1}{1+\sqrt{i\omega c_l(\omega)/\lambda r_1^2}} 
\end{equation}
It should be noted that in solving the same problem on the basis of the ordinary heat diffusion equation (\ref{e6}) one arrives at (\ref{e16}) but with $c_p$ instead of $c_l$. It is thus seen that in doing specific heat spectroscopy by effusion in a spherical geometry one obtains again the longitudinal specific heat and not the isobaric specific heat.

One can also consider the case of a soft core, $\sigma_{rr}(r_1)=0$. Although the displacement field is altered compared to the case of a hard core, the expression for the thermal impedance is still found to be given by (\ref{e16}).

The DC-limit gives the heat conductivity,
\begin{equation}
Z \rightarrow \frac{1}{4 \pi \lambda r_1} \quad \textnormal{for} \quad \omega \rightarrow 0
\end{equation}
The high-frequency limit is in concordance with the one-dimensional result,
\begin{equation}
Z \rightarrow \frac{1}{4 \pi r_1^2\sqrt{i\omega c_l(\omega)\lambda}} \quad \textnormal{for} \quad \omega \rightarrow \infty
\end{equation}
since short thermal waves cannot "see" the curvature of the sphere.
It is seen that effusion in spherical geometry in fact gives information on two properties, the heat conductivity and the heat capacity, whereas the unidirectional effusion only gives the effusivity. This is because a characteristic length scale, the radius of the heat-producing spherical surface, is involved. Effusivity in spherical geometry thus makes it possible to derive the heat capacity absolutely. However the practical usable frequency range will be more limited for a given sensitivity since the contribution from $c_l$  in (\ref{e16}) will vanish at low frequencies. At high frequency the possibility of modelling the contribution to the thermal impedance from the heat-producing device itself will also put a limit.

In real plane-plate effusion experiments the finite width of the plate gives rise to boundary effects when the heat diffusion length becomes comparable to the plate width. The deviation from the simple formula (\ref{e0}) is dependent on the ratio between these to quantities. Since a length now appears in the problem this deviation again gives the possibility of determining $\lambda$ separately. This has been addressed perturbatively \cite{bir97} on the basis of the ordinary heat diffusion equation (\ref{e6}), but not with the more exact coupled thermomechanical equations (\ref{e3}) and (\ref{e4}). Thus in fact it seems that of the two simple idealized models - the planar and the spherical - of heat effusion including the thermomechanical coupling the spherical may be the one that mostly resembles its practical realisation.  

\section{Conclusion}
These examples - the unidirectional and the spherical geometry - seem to show that it is inherently difficult to get the \textit{isobaric} specific heat directly from effusivity measurements. However another well-defined quantity, the \textit{longitudinal} specific heat can be found.

\end{document}